\documentclass{eas}
\usepackage{graphicx}
\begin{document}
\runningtitle{A. Sozzetti: Exoplanets with Gaia: Synergies in the Making}
\title{Exoplanets with Gaia: Synergies in the Making} 
\author{A. Sozzetti}\address{INAF - Osservatorio Astrofisico di Torino, 
Strada Osservatorio, 20 - 10025 Pino Torinese Italy}
%
%
\begin{abstract}
The era of high-precision astrometry has dawned upon us. The potential of Gaia $\mu$as-level precision 
in positional measuraments is about to be unleashed in the field of extrasolar planetary systems. The Gaia data hold the promise for much 
improved global characterization of planetary systems around stars of all types, ages, and chemical composition, particularly when 
synergistically combined with other indirect and direct planet detection and characterization programs. 
\end{abstract}
\maketitle
\section{Introduction}

In a recent study, Petigura et al. (2013) re-analyzed Kepler's exquisitely precise photometric data 
to determine, by extrapolation, that $5.7^{+1.7}_{-2.2}\%$ of Sun-like stars harbor an Earth-size planet 
with orbital periods of 200-400 d. In the year of the first discovery announcement by Mayor \& Queloz (1995), 
it would have been hard to believe that two decades later a touchstone in the question of life in the Universe 
such as determining whether Earth-like planets are common or rare would loom within grasp. This is but one 
example of the astonishing results obtained in the field of extrasolar planets to-date. 

Observational data on extrasolar planetary systems probe the diverse outcomes of both giant and terrestrial planet formation, 
physical, and dynamical evolution. Large survey programs, both from the ground and in space, allow to determine planetary frequencies 
for systems with different masses, sizes, orbital characteristics, and express them as a function of the properties 
of the host stars, such as mass and chemical abundance. The diverse realizations of exoplanetary properties indicate that the 
characteristiscs of our solar system are but one outcome in a continuum of possibilities. Planetary systems composed of one or more 
planets with $\approx1-3$ R$_\oplus$ orbiting within a fraction of the Earth-Sun distance appear to outnumber 
Jupiter-sized planets (Howard 2013, and references therein). The latter class, however, is the one these days 
routinely studied through atmospheric characterization measurements (Seager \& Deming 2010) . 

Today, at the banquet of the vast array of techniques for planet detection and characterization feasting on an ever increasing 
amount of outstanding quality data, astrometry remains sitting as the elephant in the room. But one year after the launch of ESA's Cornerstone 
mission Gaia, we're finally reaching the turn of the tide. The year of 2015 might be the last in which reviews appear in the 
literature, which outline what astrometry can contribute to the field of exoplanets based on simulations. The time for 
action with actual astrometric data is around the corner!

\section{Gaia and Exoplanet Science}

The state-of-the-art of astrometric techniques is currently set by the milli-arcsecond (mas) precision
achieved by the Hipparcos mission with global position measurements and by HST/FGS narrow-angle astrometric mode. 
For the purpose of astrometrically detecting planetary-mass companions in orbit around stars in the neighborhood of our Sun, 
this performance levels are not sufficient. 

Progress has been made recently with an approach based on differential astrometry of faint sources in dense fields with 8-meter class ground-based 
telescopes. Sahlmann et al. (2013) have demonstrated long-term 200 micro-arcsecond ($\mu$as)-level precision with VLT/FORS2 on a sample of brown dwarfs, 
and announced the astrometric discovery of an ultra-cool low-mass binary (DE0823-49; $M_1=78.4$ $M_\mathrm{J}$, $M_2=28.5$ $M_\mathrm{J}$)
 with the lowest mass ratio ($q=M_2/M_1=0.36$) of known very low-mass binaries with characterised orbits.

However, for astrometry to begin contributing significantly to the fast-developing field of exoplanet
astrophysics, a quantum leap of at least one additional order of magnitude in positional measurements precision must be obtained. 

\subsection{The Challenge}

\begin{figure}
\centering
$\begin{array}{cc}
\includegraphics[width=0.45\textwidth]{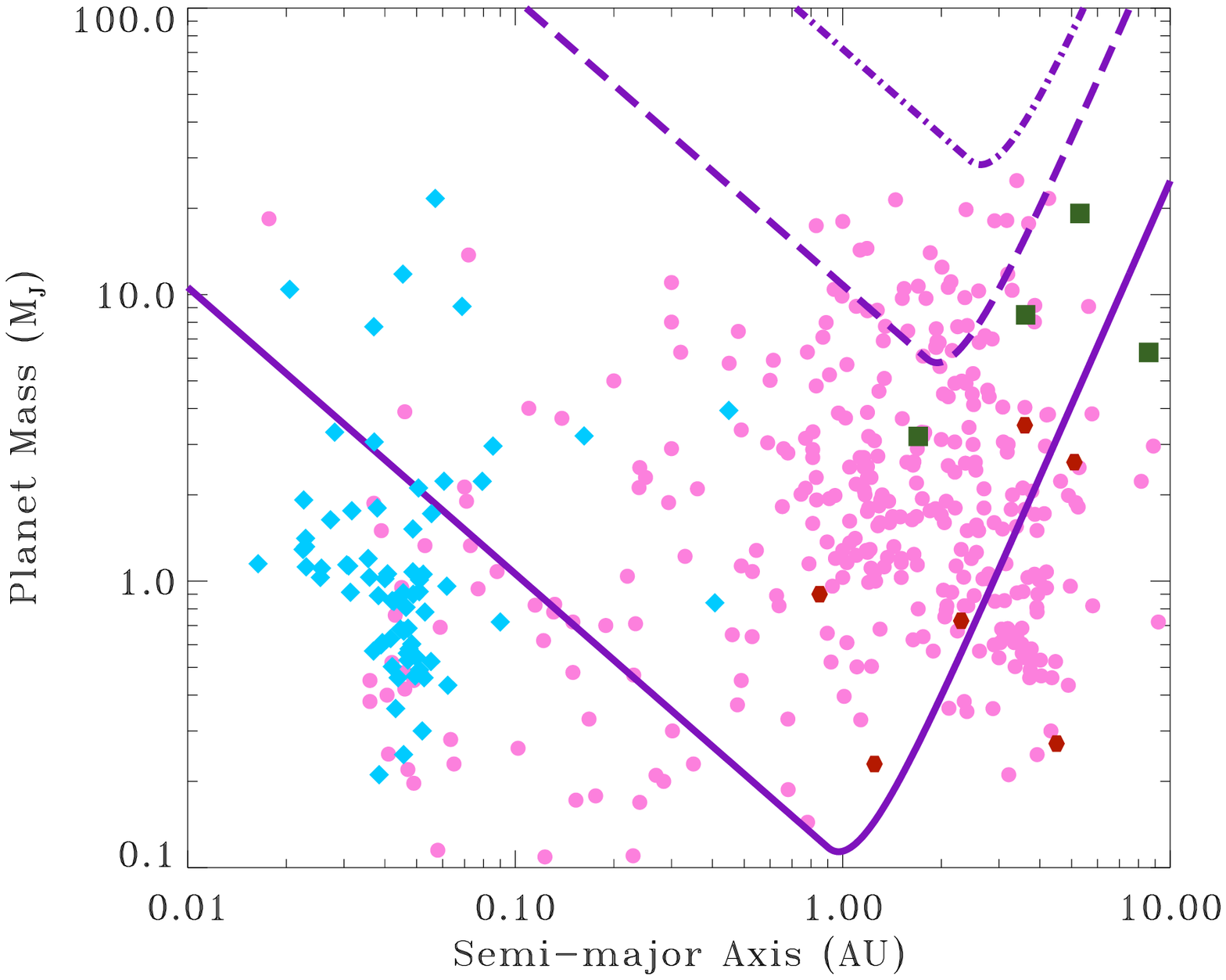} & 
\includegraphics[width=0.50\textwidth]{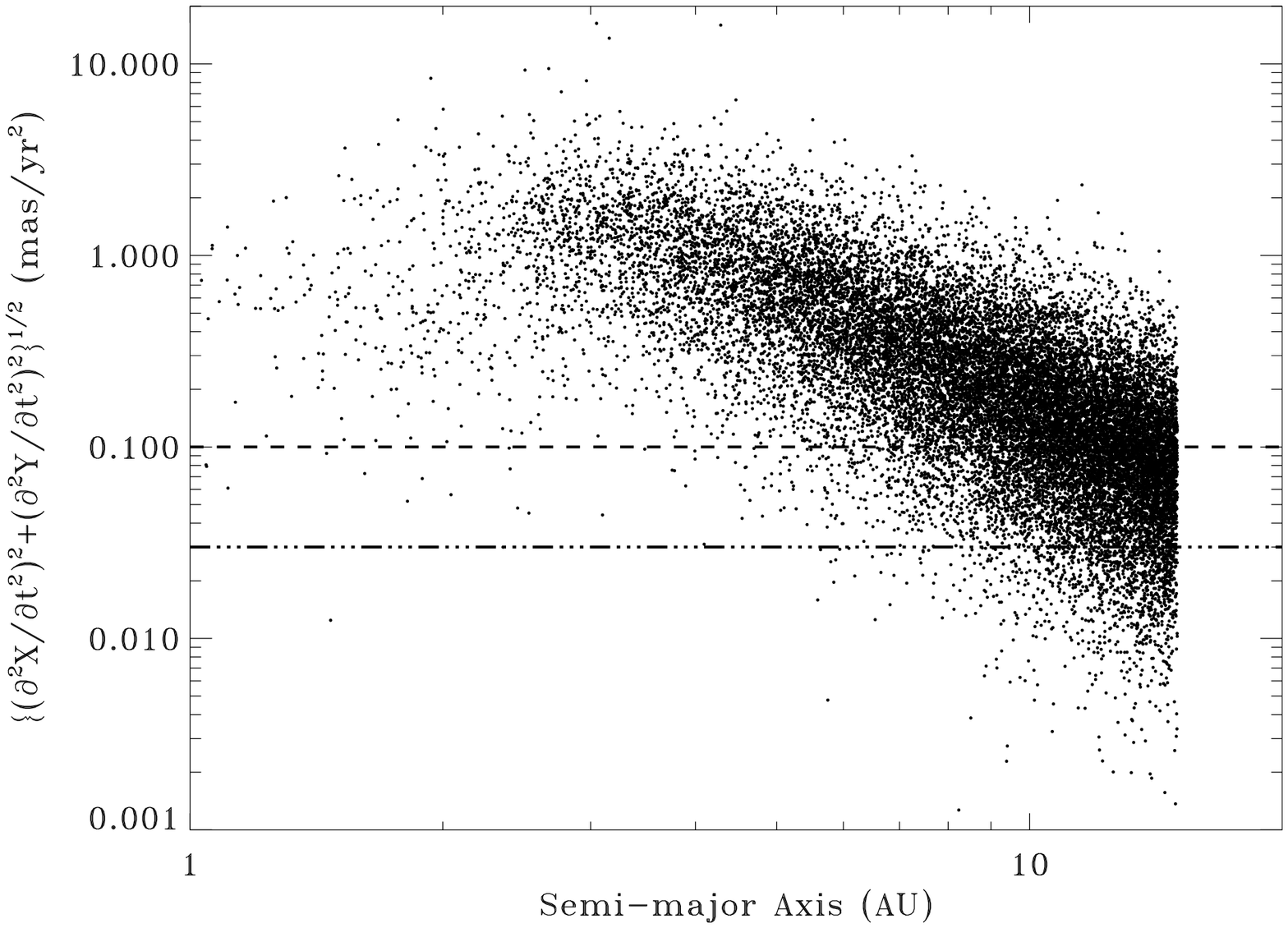} \\
\end{array} $
\caption{{\it Left:} Gaia discovery space of brown dwarf companions to stars and low-mass companions to brown dwarfs (purple curves). 
Detectability curves are defined on the basis of a 3-$\sigma_\mathrm{A}$ criterion for signal detection (see Sozzetti 2010 for details).
The upper dashed-dotted and center dashed curves are for Gaia astrometry with $\sigma_\mathrm{A} = 120$ $\mu$as,
assuming a 0.8-$M_\odot$ primary at 300 pc and for $\sigma_\mathrm{A} = 400$ $\mu$as, assuming a 0.2-$M_\odot$ primary at 30 pc, respectively. 
The lower solid curve is for $\sigma_\mathrm{A} = 500$ $\mu$as, assuming a 0.050-$M_\odot$ primary at 2 pc (appropriate for Luhman 16A). 
The survey duration is set to 5 yr. The pink filled circles indicate a representative samples of Doppler-detected exoplanets. Transiting systems
are shown as light-blue filled diamonds. Red hexagons are planets detected by microlensing. Planets detected with the timing 
technique are also shown as green squares. {\it Right:} Accelerations in the stellar motion induced by $1-70$ $M_J$ orbiting companions at orbital 
separations $a<15$ au detected by SPHERE around a sample of $>400$ targets of the GTO program with $V<12$ mag and $d< 50$ pc. 
Dashed and dashed-dotted lines indicate 3-$\sigma_\mathrm{A}$ detection limits with Gaia at mid-mission (2.5 yr) and at mission end (5 yr). Only 5\% of the 
accelerations in Gaia astrometry go undetected at the end of the mission, with a high degree of completeness (99\%) for $a<7$ au.}
\label{fig1}
\end{figure}

Gaia, now operating at L2 since more than a year after its successful launch in December 2013, is the first experiment set to
demonstrate single-epoch measurement accuracies $\sigma_\mathrm{A}\approx20$ $\mu$as for bright stars (see the reviews on the mission status and 
performance by Prusti and de Bruijne, this volume). Gaia global astrometry will finally enable secure astrometric detections of
planetary-mass companions around nearby solar-type stars. However, this is no easy task. There is in fact a variety of technical problems associated
with the modeling of the astrometric signatures of planetary systems that must be carefully dealt with (e.g., Sozzetti 2005). It is 
worth underlining that fitting astrometric orbit of exoplanets (particularly in the case of multiple companions) involves the adjustment of a large number of 
parameters, many of which nonlinear. The assessment of their reliability and robustness (including meaningful error estimates on the fitted quantities) 
will be a nontrivial task, particularly in the limit of astrometric signals comparable in size to Gaia's single-measurement uncertainties and/or limited 
redundancy in the number of observations with respect to the model parameters (see, e.g., Sozzetti 2012). 
For these reasons, within the pipeline of Coordination Unit 4 (object processing) of the Gaia Data Processing and Analysis Consortium (DPAC), 
in charge of the scientific processing of the Gaia data and production of the final Gaia catalogue to be released sometime in 2021, 
a Development Unit (DU437) has been specifically devoted to the modelling of the astrometric signals produced by planetary systems. 
The DU is composed of several tasks, which implement multiple robust procedures for (single and multiple) astrometric orbit fitting (such as Markov Chain Monte Carlo 
and genetic algorithms) and the determination of the degree of dynamical stability of multiple-component systems. 

\subsection{Discovery Potential}



The size of the astrometric perturbation induced on the primary by an orbiting planet (the astrometric signature) corresponds to the 
semi-major axis of the orbit of the primary around the barycenter of the system scaled by the distance to the observer: 
$\alpha=(M_p/M_\star)\times(a_p/d)$. With $a_p$ in au, $d$ in pc, and $M_p$ and $M_\star$ in M$_\odot$, then $\alpha$ is evaluated in arcsec. 
At 10 pc, a $M_p=1$ $M_\oplus$ planet at the center of the circumstellar Habitable Zone\footnote{In its classical definition, this is the region around a star within which 
planetary-mass objects with sufficient atmospheric pressure can support liquid water at their surfaces (see, e.g,., Kopparapu et al. 2014, and references therein.)} 
($a_p=1$ au) of a solar-like star ($M_\star=1$ $M_\odot$) induces an orbital motion with an amplitude $\alpha=0.3$ $\mu$as. 
It thus clear how for Gaia, even assuming its best-achievable astrometric precision 
of a few tens of $\mu$as, the discovery of terrestrial planets lies beyond the realm of its capabilities.  

The sensitivity of Gaia astrometry to (single and multiple) giant planetary companions at intermediate separations around bright, nearby, F-G-K-type dwarfs 
has been quantified in the past by Lattanzi et al. (2000), Sozzetti et al. (2001), and Casertano et al. (2008). More recently, 
Sozzetti et al. (2014) and Perryman et al. (2014) have revisited those early estimates based on improved (pre-commissioning) knowledge of the 
astrometric error budget and extending the studies to encompass a wider range of primary spectral types and limiting target magnitudes (down to $G=20$ mag). 
The global figures on which all the above works converge speak of several thousands (possibly 10-20$\times10^4$) astrometrically detectable giant planets 
at separations between typically 0.5 au and $4-5$ au from their parent stars. The overall all-sky reservoir of stars around which Gaia will be sensitive 
to planetary-mass companions thus largely exceeds $10^6$. 
Finally, based on recent representations of the design of Gaia and its expected photometric performance, Dzigan \& Zucker (2012) showed that 
Gaia should provide a sample of maybe $\approx10^3$ transiting hot Jupiters around main-sequence solar-type stars. 

\subsection{Specific Object Classes} 

Gaia, being fundamentally unbiased in its all-sky magnitude-limited survey, will monitor astrometrically stellar and substellar objects, 
with no discrimination for spectral type, age, evolutionary, and multiplicity status. Studies are now beginning to gauge 
Gaia sensitivity to planets around not your typical F-G-K-M primary. 

Sozzetti (2014) has recently shown that Gaia will not only monitor astrometrically millions of main-sequence stars with sufficient sensitivity 
to brown dwarf companions within a few au from their host stars, but that thousands of detected ultra-cool dwarfs in the backyard of the Sun will have direct 
distance estimates from Gaia. For these, Gaia astrometry might be of sufficient precision to reveal any orbiting companions with masses even below 1 M$_J$ (see Figure 1).

Silvotti et al. (2014) have reported on exploratory experiments in which completeness limits in the astrometric detection of massive (5-15 $M_\mathrm{J}$) 
planets and mid-range ($\sim50$ $M_\mathrm{J}$) brown dwarfs at intermediate separations from white dwarf primaries with Gaia extend out to 20-40 pc and 70 pc, respectively. 

Finally, Sahlmann et al. (2015) have started tackling the problem of astrometric planet detection in binary stellar systems. They found that Gaia might 
discover hundreds of circumbinary giant planets in systems with F-G-K dwarf primaries within 200 pc of the Sun, assuming similar giant planet mass distribution and 
occurrence rates for tight binaries and single stars. If on the other hand all circumbinary gas giants have masses lower than 2 $M_\mathrm{J}$ 
(as inferred based on results from the Kepler mission), Gaia detections might be reduced to a few.

\subsection{The Gaia Legacy}

As a direct consequence of its unbiased census of thousands of planetary systems, the actual impact of Gaia measurements in exoplanets science will be broad and structured. 
For example, the Gaia data will: {\bf a)} allow to test the fine structure of 
giant planet parameters distributions and frequencies (including the transition region between giant planets and brown dwarfs), and investigate their changes as a function of 
stellar mass, metallicity, and age with unprecedented resolution; {\bf b)} help crucially test theoretical models of the formation and migration of gas giant planets, 
and their impact on the formation scenarios for terrestrial planets (see Johansen, this volume); {\bf c)} achieve key improvements in our comprehension of important aspects of
the formation and dynamical evolution of multiple-planet systems via direct measurements of their relative orbital arrangement; {\bf d)} provide the first-ever statistically 
robust estimates of giant planet frequencies around ultra-cool dwarfs and around stars in the final evolutionary states (e.g., white dwarfs), as fundamental testing ground 
for the hypothesis that planet formation processes may not stop around sub-stellar mass primaries and providing crucial observational support for distinguishing between 
scenarios of post-main-sequence planetary systems evolution and second-generation planet formation processes.

\section{The Gaia Treasure Trove of Synergies}

The broad range of applications to exoplanets science is such that Gaia data can be 
seen as an ideal complement to (and in synergy with) many ongoing and future observing programs devoted 
to the indirect and direct detection and characterization of planetary systems, both from the ground and in space. 

\subsection{Synergies with Direct Imaging Programs}

Gaia data on long-period planets will inform direct imaging surveys and spectroscopic characterization projects 
about the epoch and location of maximum brightness of (primarily non transiting) exoplanets, in order to estimate their optimal visibility. 
Work in progress (Sozzetti et al. in preparation) is focusing on the effectiveness of the combination of SPHERE/VLT direct detections of 
wide-separation giant planets with Gaia determinations of accelerations in the stellar motion due to the orbiting companions for improved 
constraints on the orbital architecture and mass, thereby helping in the modeling and interpretation of giant planets' phase functions and light curves 
(see Figure 1). 

\subsection{Synergies with Ground-Based Transit Programs}

Recent findings (Sozzetti et al. 2014; Perryman et al. 2014) indicate that Gaia might identify tens if not hundreds of potentially transiting 
intermediate-separation giant planets (i.e. with astrometric orbits compatible with $i=90^\circ$, within a few degrees). Such systems, in which 
planets might transit and/or be occulted by their relatively bright primaries would then become very interesting targets for follow-up photometry, 
to ascertain whether the prediction is verified or not. The possibility to study a sample of transiting cold (i.e., long-period) 
giant planets is certainly intriguing, for systematic comparison with their strongly irradiated, short-period counterparts in terms of mass-radius 
relationship and atmospheric characterization. Dedicated photometric follow-up efforts would also help discriminating among and 
improving the characterization of the many transiting hot Jupiter candidates expected from Gaia photometry (Dzigan \& Zucker 2012)

\subsection{Synergies with Space-Borne Transit Programs}

The availability of very accurate direct distance measurements (a few percent) to all bright stars in the sky, 
starting with Gaia early data releases in 2017, will be a critically needed contribution to the definition of the input catalogues for space-borne 
photometric transit surveys such as those that will be carried out by TESS (Ricker et al. 2014) and PLATO (Rauer et al. 2014). 
It will be in fact possible to define stellar samples of nearby 
solar-type main-sequence stars with negligible contamination from distant giant stars. Gaia parallaxes will also be instrumental in the improved 
determination of the fundamental physical properties (mass radius) of the hosts of (candidate and confirmed) transiting planets from Kepler, K2, and CHEOPS 
(and ground-based transit surveys such as WASP and HAT), thus allowing to improve the measurements of the planetary parameters themselves. 

\subsection{Synergies with Doppler Surveys}

There is a very strong, two-fold synergy potential between Gaia astrometry and high-precision Doppler measurements gathered with ongoing, upcoming and planned 
instrumentation operating at both visible (e.g., HARPS, HARPS-N, ESPRESSO, HIRES/E-ELT) and infrared wavelengths (e.g., GIANO, SPIROU, CARMENES). 
First, the combination of RVs of all bright ($V<13$ mag), nearby ($d< 200$ pc) stars hosting planets and Gaia astrometric data will allow to 
a) characterize planetary systems across orders of magnitude in mass and orbital separation, b) improve studies of the dynamical evolution of multiple 
systems with giant planets, including meaningful coplanarity analyses. Second, high-resolution, high-precision Doppler programs will cherry-pick on 
Gaia astrometric detections with the three-fold aim of 1) improving the phase sampling of the astrometric orbits determined by Gaia, b) extending the time 
baseline of the observations (to put stringent constraints on or actually characterize long-period companions), and c) search for additional, low-mass and/or 
short-period components which might have been missed by Gaia due to lack of sensitivity. 

\section{Summary}

Much as its predecessor Hipparcos, Gaia is bound to set the standards in high-precision astrometry for the next decade or two. 
The largest compilation of new, high-accuracy astrometric orbits of giant planets, unbiased across all spectral types, and exquisitely precise parallaxes 
will define Gaia's role in the exoplanet arena. Its huge synergy potential with ongoing and planned exoplanet detection and (atmospheric) characterization programs, 
both from the ground and in space, wil allow Gaia to crucially contribute to many an aspect of the formation, physical and dynamical evolution of planetary systems. 

\section*{Acknowledgements}

I am especially grateful to the Conference organizers for giving me the opportunity to participate to this scientifically 
vibrant and smoothly run event. This work has been funded in part by ASI under contract to INAF I/058/10/0 (Gaia Mission: The Italian Participation to DPAC).


\end{document}